\documentclass[twocolumn,preprintnumbers,prd]{revtex4}

\usepackage{graphicx}
\usepackage{xcolor}

\usepackage[hypertex]{hyperref}
\usepackage{amsmath, amssymb}

\newcommand{\lsim}{\lesssim}
\newcommand{\ord}[1]{\mathcal{O}{(#1)}}

\newcommand{\gsim}{\gtrsim}
\newcommand{\eq}[1]{Eq.~(\ref{#1})}
\newcommand{\beq}{\begin{equation}}
\newcommand{\eeq}{\end{equation}}

\newcommand{\eps}{\varepsilon}

\newcommand{\res}{R_{\rm ES}}
\newcommand{\rau}{R_{\rm AU}}
\newcommand{\mat}[2][ccccc]{\left( \begin{array}{#1} #2\\ \end{array}\right)}

\begin{document}

\pagestyle{plain}

\title{Long-Range Lepton Flavor Interactions and Neutrino Oscillations}

\author{Hooman Davoudiasl\footnote{email: hooman@bnl.gov}}

\author{Hye-Sung Lee\footnote{email: hlee@bnl.gov}}

\author{William J. Marciano\footnote{email: marciano@bnl.gov}}

\affiliation{Department of Physics, Brookhaven National Laboratory, Upton, NY 11973, USA}


\begin{abstract}
Recent results from the MINOS accelerator neutrino experiment suggest a possible difference between $\nu_\mu$ and
${\bar \nu_\mu}$ disappearance oscillation parameters, which one may ascribe to a
new long-distance potential acting on neutrinos.  As a specific example, we consider a model
with gauged $B-L_e-2 L_\tau$ number that contains an extremely light new vector boson $m_{Z'} < 10^{-18}$~eV and
extraordinarily weak coupling $\alpha' \lsim 10^{-52}$ 
(or larger $m_{Z'}$ if cosmology bounds on neutrino decay apply).  
In that case, differences between $\nu_\mu \to \nu_\tau$
and ${\bar \nu_\mu}\to {\bar \nu_\tau}$ oscillations can result from a long-range
potential due to neutrons in the Earth and the Sun that distinguishes
$\nu_\mu$ and $\nu_\tau$ on Earth, with a potential difference of
$\sim 6\times 10^{-14}$~eV, and changes sign
for anti-neutrinos.  We show that existing solar, reactor, accelerator, and
atmospheric neutrino oscillation constraints can be largely
accommodated for values of parameters that help explain the possible MINOS
anomaly by this new physics, although there is some tension with atmospheric constraints.
A long-range interaction, consistent with current bounds,
could have very pronounced effects on atmospheric neutrino disappearance in the 15 - 40~GeV
range that will be studied with the IceCube DeepCore array, currently in operation, and can have a significant effect on
future high-precision long-baseline oscillation experiments that aim for $\pm 1\%$ sensitivity,
in $\nu_\mu$ and $\bar\nu_\mu$ disappearance, separately.   Together, these
experiments can extend the reach for new long-distance effects well beyond current bounds and test their
relevance to the aforementioned MINOS anomaly.  We also point out
that long-range potentials originating from the Sun could lead to annual modulations
of neutrino data at the percent level, due to the variation of the Earth-Sun distance.  A similar phenomenology is shown 
to apply to other potential new gauge symmetries such as $L - 3 L_\tau$ and $B - 3 L_\tau$.
\end{abstract}

\maketitle

\section{Introduction}
\label{introduction}
Neutrino flavor oscillation experiments have provided some of the most direct and robust
indications of physics beyond the Standard Model (SM).  Solar, atmospheric,
reactor, and accelerator data all point to the conclusion that at least 2 active neutrinos
have tiny but non-zero masses of up to order $0.1$~eV,
whose generation requires extending the SM.  We refer the interested reader to
Refs.~\cite{GonzalezGarcia:2007ib,Nakamura:2010zzi} for a review of the extensive literature on neutrino oscillation physics.
Given the smallness of neutrino mass differences, even minute perturbations
to the time evolution of flavor eigenstates, caused by feeble differences of interactions of neutrinos with
background sources, can produce measurable departures from vacuum oscillations.
For example, these effects can be caused by the short-distance electroweak
interactions of neutrinos with solar or terrestrial electrons, referred to as the
Mikheev-Smirnov-Wolfenstein (MSW) effect \cite{Wolfenstein:1977ue,Mikheev:1986gs}.  The
sensitivity of neutrino oscillations to such small effects makes them a good probe of new
physics that violates $\nu_e$-$\nu_\mu$-$\nu_\tau$ universality \cite{Botella:1986wy}.
Hence it is interesting to look for unexpected effects in neutrino data.

Recently, measurements at the MINOS experiment \cite{MINOS} have resulted in different
inferred values for differences of squared masses and mixing angles
\beq
| \Delta m_{23}^2 | = 2.35^{+0.11}_{-0.08} \times 10^{-3} ~{\rm eV}^2 ; \;  \sin^2(2\theta_{23}) =1.00
\label{minosnu}
\eeq
[where $\sin^2(2\theta_{23}) =1.00$ is the best fit
value, while $\sin^2(2\theta_{23}) > 0.91$ at 90\% confidence level] and
\beq
| \Delta {\bar m_{23}}^2 | = 3.36^{+0.45}_{-0.40} \times 10^{-3} ~{\rm eV}^2 ; \;
\sin^2(2{\bar \theta}_{23}) = 0.86\pm 0.11
\label{minosnubar}
\eeq
in $\nu_\mu$ and ${\bar \nu}_\mu$ disappearance, respectively.
The above MINOS results have revived some interest in
long-range interactions (LRIs)
\cite{Heeck:2010pg} that can cause disparities between neutrinos and
anti-neutrinos.  For other related works on the MINOS anomaly, see, for instance,
Refs.~\cite{Engelhardt:2010dx,Mann:2010jz,Kopp:2010qt}.

The possibility of new long-range forces was discussed
in the pioneering work of Ref.~\cite{Lee:1955vk}, and subsequently
considered as an alternative way to explain apparent CP violating effects
in $K$ meson decays \cite{Bell:1964ff,Bernstein:1964hh}.  Note that the disparity in
the oscillation parameters for neutrinos and anti-neutrinos,
as suggested by the MINOS results (\ref{minosnu}) and (\ref{minosnubar}), can be
ascribed to an apparent violation of CPT \cite{Barenboim:2009ts}.  However, in what
follows we will assume that CPT is conserved in vacuo and consider the
possibility that the MINOS result could be a hint of a new LRI.
E\"{o}tv\"{o}s-type \cite{Eotvos:1922pb} tests of gravity place
stringent bounds on these interactions \cite{Lee:1955vk},
constraining their ``fine structure constant"
$\alpha' \leq 10^{-49}$ (electron coupling) and $\alpha' \leq 10^{-47}$
(nucleon coupling) \cite{Okun:1995dn,Dolgov:1999gk}.  This
suggests an astrophysical source with a large number of particles is needed, for
sizable long-range effects.  Long-range interactions  in neutrino
oscillations were considered in 
Refs.~\cite{Grifols:2003gy,JM,GonzalezGarcia:2006vp, Bandyopadhyay:2006uh};  
see also Ref.~\cite{Samanta:2010zh}.  
The long-range vector interaction yields equal and opposite potentials
for leptons and anti-leptons.  This can then manifest itself as a difference in the properties of
neutrinos and anti-neutrinos in terrestrial oscillation experiments, caused by the collective
effect of particles in the Sun and the Earth charged under a new $U(1)'$ gauge symmetry.
The corresponding effective fine structure constant must be extremely small,
$\lsim \ord{10^{-49}-10^{-47}}$, as required
by precision tests of gravity \cite{gravtest}.  For comparison, note that  the effective
gravitational coupling between two protons is of order $\alpha_g \sim G_N \, m_p^2 \sim 10^{-38}$,
where $G_N$ is Newton's constant and $m_p$ is the proton mass.
Here, it is assumed that the associated $Z'$
vector boson has a mass $m_{Z'} < 1/R_{\rm AU}$,
where $R_{\rm AU} = 1~{\rm AU} \simeq 1.50\times 10^8 {\rm ~ km} \sim 10^{18}$~eV$^{-1}$ is
the mean Earth-Sun distance.  (Later, we will limit our discussion to values of $m_{Z'}$ that are not far below
$10^{-18}$~eV, in order to exclude contributions from the rest of the Galaxy.)

Before going further, we would like to make a few comments regarding the results
(\ref{minosnu}) and (\ref{minosnubar}).  First, the suggested MINOS anomaly
is not at statistically significant levels, being at most a 2-sigma effect.
In addition, the available atmospheric data from MINOS yield the best fit \cite{MINOSatm} (2-state mixing)
\beq
| \Delta m^2 | - | \Delta {\bar m}^2 |= 0.4^{+2.5}_{-1.2} \times 10^{-3} ~{\rm eV}^2
\label{minos_atm}
\eeq
which does not support the above accelerator results, and, while also statistically limited, very mildly prefers an
opposite sign for the effect.  Taken together, the above considerations do not provide a strong
case for invoking new physics.  Nevertheless, we find the MINOS accelerator data sufficiently intriguing
to motivate an examination of the prospects for probing long-range leptonic forces at current
and future experiments, as detailed  below.

In what follows, we will discuss the possibility of attributing the aforementioned MINOS
anomaly to a LRI potential, generated by the neutrons in the Earth and the Sun.
We will show that the existing bounds from neutrino oscillation data
do not exclude such an interpretation.  We use our approximate fit
as a benchmark for potentially interesting values of parameters
and estimate the reach of current and future experiments for the new LRI.
We find that the IceCube DeepCore array \cite{Wiebusch:2009jf}, which
is currently in operation,
can provide an excellent probe
of the benchmark model parameters and reach well-beyond them.
We point out that long-range potentials generated by solar particles
will inevitably lead to {\it annual modulation} of neutrino
oscillations at Earth, due to the variation of the Earth-Sun distance.  The large
event sample expected at DeepCore seems sufficient to uncover a possible effect
at the 1\% level, statistically.  Observation of such
modulations can provide a distinct clue as to the solar contribution to the
LRI and set a lower bound on its range.  We will also consider long-baseline
experiments, such as those envisioned for the
Deep Underground Science and Engineering Laboratory (DUSEL) \cite{Raby:2008pd},
to discover or constrain various effects of the LRI.  We find that the expected
capabilities of these experiments  would allow them to probe the difference
between the oscillation parameters of neutrinos and anti-neutrinos, induced by LRIs,
which is a key signal for this type of new physics.  Although we concentrate on the case of a 
potential generated by neutrons, our results and analysis carry over to other interesting 
scenarios where, for example, electrons or the total baryon number of the Sun and Earth 
may be responsible for the LRI.  

We will next briefly present the basic setup and formalism used in our work.
Section~\ref{phenomenology} will contain our analysis and results.  Our concluding
remarks will be presented in Section~\ref{conclusions}.

\section{Formalism}
\label{formalism}
Let us consider the addition of a general anomaly-free $U(1)'$ gauge quantum number 
(for vectorial representations) to the SM \cite{Ma:1997nq,Lee:2010hf} 
\beq
{\cal Q}= a_0 (B-L) + a_1 (L_e-L_\mu) + a_2 (L_e-L_\tau) + a_3 (L_\mu-L_\tau),
\label{Q}
\eeq
where $B$ and $L$ are baryon and lepton numbers,
respectively, while $L_\ell, \ell=e,\mu, \tau$
are lepton flavor numbers, and $a_i$, $i=0,1,2,3$,
are arbitrary constants.  For our primary example, we will set $a_1=a_2=0$ and
for definiteness take $a_0=a_3=1$.  However, any values of $a_0$ and $a_3$ will lead to
the same neutrino oscillation phenomenology for a fixed coupling between
$B-L$ and $L_\mu-L_\tau$.  In this combination of quantum numbers,
\beq
{\cal Q} = (B-L) + (L_\mu-L_\tau) = B-L_e-2 L_\tau,
\label{Q}
\eeq
$(B-L)$ is associated with the source of the new potential, while
$(L_\mu-L_\tau)$ provides a contribution to the relevant neutrino oscillation $\nu_\mu - \nu_\tau$.
Our choice for ${\cal Q}$ in \eq{Q}, as we will later argue,
is less constrained by experiments than the previously studied $L_e-L_{\ell}$, $\ell=\mu, \tau$,
cases.  It also follows that the LRI potential due to $B-L$ that we consider
is generated by the total neutron number, since the contributions of
electrons and protons cancel.

Our charge assignment provides a simple way of achieving the effective
coupling in Ref.~\cite{Heeck:2010pg}, where the microscopic origin of
the requisite interactions is a mixing between a $Z'$ associated with
$L_\mu - L_\tau$ number and the $Z$ boson of the SM.  In principle, one could also
imagine a mixing between two $Z'$ states associated with, say, $B-L$ and $L_\mu - L_\tau$,
where an appropriate choice of mixing parameters will yield the effective scenario adopted here.
Given that our main purpose in this work is to elucidate the relevant phenomenology, without
reference to a particular underlying theoretical context, our choice of the gauged quantum
number captures all the relevant key features for our analysis,
while avoiding unnecessary complications.
Note that as long as one of the anomaly-free quantum numbers
is carried by a main constituent of solar or terrestrial matter,
with the other lepton flavor number differences, one can build models that
result in qualitatively similar effects.  Indeed, we later consider models with 
gauged $L-3L_\tau$ and $B-3L_\tau$ that exhibit essentially the same phenomenology 
as that of $B-L_e-2L_\tau$, but with even smaller gauge couplings.  

The range of the interaction corresponding to charge ${\cal Q}$ is
determined by the mass $m_{Z'}$ of the force carrier $Z'$.  Since we are
interested in the effect of a large number of particles, we assume that
$m_{Z'} \lsim 10^{-18}$~eV so that the neutrons both in the Earth and the Sun can contribute.  
We will not consider $m_{Z'}\ll 10^{-18}$~eV so that our assumed
LRI does not extend far beyond the solar system and
the contribution of the rest of the galaxy can be ignored \cite{footnote}.
The resulting potential
felt by neutrinos on the Earth is then given by
\begin{eqnarray}
V_n &= &\alpha' \left(\frac{N_n^\oplus}{R_\oplus} +
\frac{N_n^\odot}{R_{\rm ES}}\right)
= 2.24\times 10^{-12}~{\rm eV} \nonumber\\
&\times& \left(\frac{\alpha'}{10^{-50}}\right)
\left[0.25 + \left(\frac{\rau}{\res}\right)\right],
\label{Vn}
\end{eqnarray}
using the estimated solar neutron fraction $Y_n^\odot = 1/7$ ({\it i.e.} $N_p^\odot/N_n^\odot \simeq 6$,
where $N_p^\odot$ is the number of protons in the Sun) and $Y_n^\oplus = 1/2$ for the neutron fraction in the Earth.
In Eq.~(\ref{Vn}), $N_n^\oplus = 1.78 \times 10^{51}$ and $N_n^\odot = 1.70 \times 10^{56}$
are numbers of neutrons in the Earth and the Sun, respectively,
$R_\oplus = 6.4\times 10^3$~km is the Earth's radius, and
$\res$ is the variable distance of the Earth from the Sun.
We note that $\res$ attains
its maximum value $\res^a \simeq 1.52 \times 10^8$~km at the aphelion ($\sim$  July 4)
and its minimum value $\res^p\simeq 1.47 \times 10^8$~km at the perihelion ($\sim$ January 4).

The ratio of the potential $V^\oplus_n$
at the Earth's surface from its neutrons to $V_n^\odot $ from solar neutrons is given by
\beq
\frac{V_n^\oplus}{V_n^\odot}
\approx \frac{1}{4}\,.
\label{Vratio}
\eeq
Thus, the contribution of the Earth-generated potential is sub-dominant but not negligible.
Note that if electrons are the source of the long-range
potential, one can show that the solar contribution is roughly 24 times larger than
that generated by  the Earth \cite{JM,Bandyopadhyay:2006uh}.

As discussed in Refs.~\cite{Botella:1986wy,Grifols:2003gy,JM}, the $\nu_\mu$
survival probability in the 2 flavor $\nu_\mu - \nu_\tau$
oscillation [for $\sin^2(2\theta_{13})\simeq 0$, $\Delta m_{12}^2 \simeq 0$] is given by
\beq
\tilde{P}_{\mu\mu} = 1 - \sin^2(2\tilde{\theta}_{23})
\sin^2\left(\frac{\Delta \tilde{m}^2_{23} L}{4 E_\nu}\right),
\label{Pmumu}
\eeq
where
\beq
\Delta \tilde{m}^2_{23} = \Delta m^2_{23}
\sqrt{[\xi - \cos(2\theta_{23})]^2 + \sin^2(2\theta_{23})}
\label{m23til}
\eeq
and
\beq
\sin^2(2\tilde{\theta}_{23})  = \frac{\sin^2(2\theta_{23})}
{[\xi - \cos(2\theta_{23})]^2 + \sin^2(2\theta_{23})}.
\label{sin23til}
\eeq
Here, the symbols that are tilde-free denote vacuum quantities, and
\beq
\xi \equiv - \frac{2 W_\tau \, E_\nu}{\Delta m^2_{23}},
\label{xi}
\eeq
with $W_\tau = {\cal Q}_\tau V_n$ the potential energy for $\nu_\tau$;
${\cal Q}_\tau = -2$ is the charge of $\nu_\tau$, in our model.
One can obtain the $\bar{\nu}_\mu$ survival probability from the above expressions
by $\xi \to -\xi$, and there is a degeneracy if both $\Delta m^2_{23}$ and $\cos(2\theta_{23})$
change sign.  Note that if $\sin^2 (2 \theta_{23}) = 1$ the formalism yields the
same results for $\nu$ and $\bar\nu$.
It should also be noted that the $\Delta \tilde{m}^2_{23}$ and
$\sin^2 (2 \tilde{\theta}_{23})$ are energy-dependent for $\alpha'\neq0$ and
deviations from the vacuum values tend to increase with energy.

The $\sin^2 (2\tilde\theta_{23})$ and $\Delta \tilde m^2_{23}$ measure the depth and location ($E_\nu \approx \Delta \tilde m_{23}^2 L / 2\pi$) of the first oscillation minimum in the survival probability $\tilde P_{\mu\mu}$ versus energy $E_\nu$.
With $X \equiv |\xi - \cos(2\theta_{23})|$, in the limit of $X \simeq 0$ (resonance condition), we have $\sin^2 (2\tilde\theta_{23}) \simeq 1$ and $\Delta \tilde m^2_{23} \simeq \Delta m_{23}^2 \sin (2\theta_{23})$.
As $X$ increases, $\sin^2 (2\tilde\theta_{23})$ decreases and $\Delta \tilde m^2_{23}$ increases.
For a negligibly small $\cos(2\theta_{23})$ where $X \simeq | \xi | = | -2 {\cal Q}_\tau V_n E_\nu / \Delta m^2_{23} |$, $\sin^2 (2\tilde\theta_{23})$ decreases and $\Delta \tilde m^2_{23}$ increases with $E_\nu$ for both $\nu$ and $\bar \nu$, for a given $\Delta m_{23}^2$.
For a sizable $\cos(2\theta_{23})$, $X$ may increase/decrease with $E_\nu$ depending on the relative sign of the two terms in $X$ as long as $|\xi| < |\cos(2\theta_{23})|$.
This means deviations from the standard oscillations are different for $\nu$ and $\bar\nu$.
If a new potential (or in general $X$) is too large, $\sin^2 (2\tilde\theta_{23}) \approx 0$ and the oscillation would be quenched.

\section{Phenomenology}
\label{phenomenology}
\begin{figure*}[tb]
\begin{center}
\includegraphics[width=0.32\textwidth]
{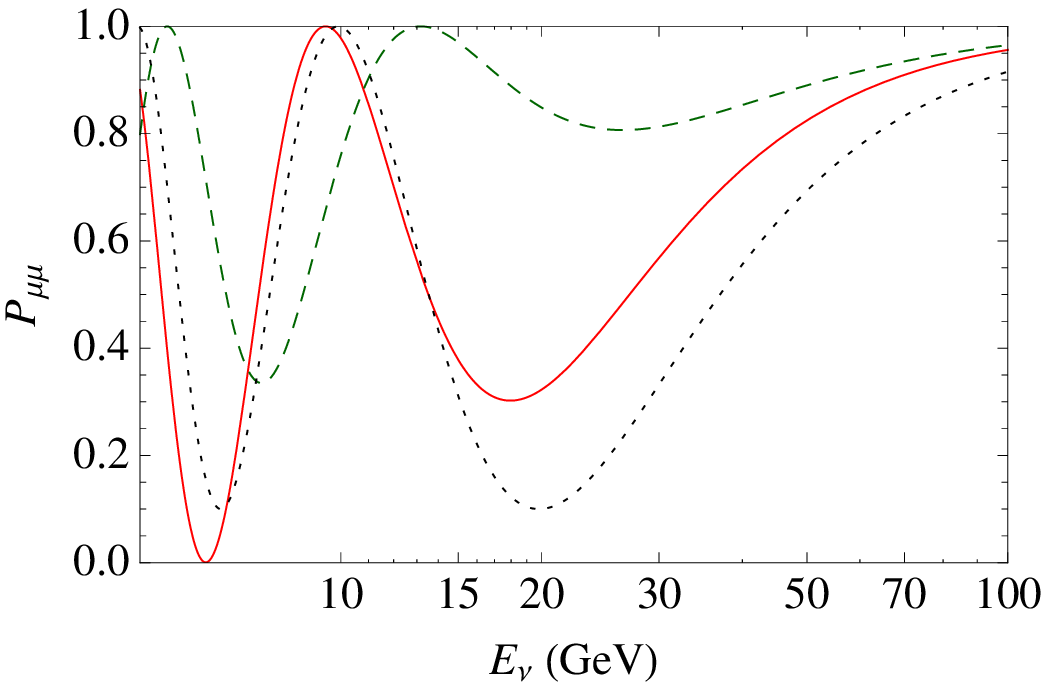} ~
\includegraphics[width=0.32\textwidth]
{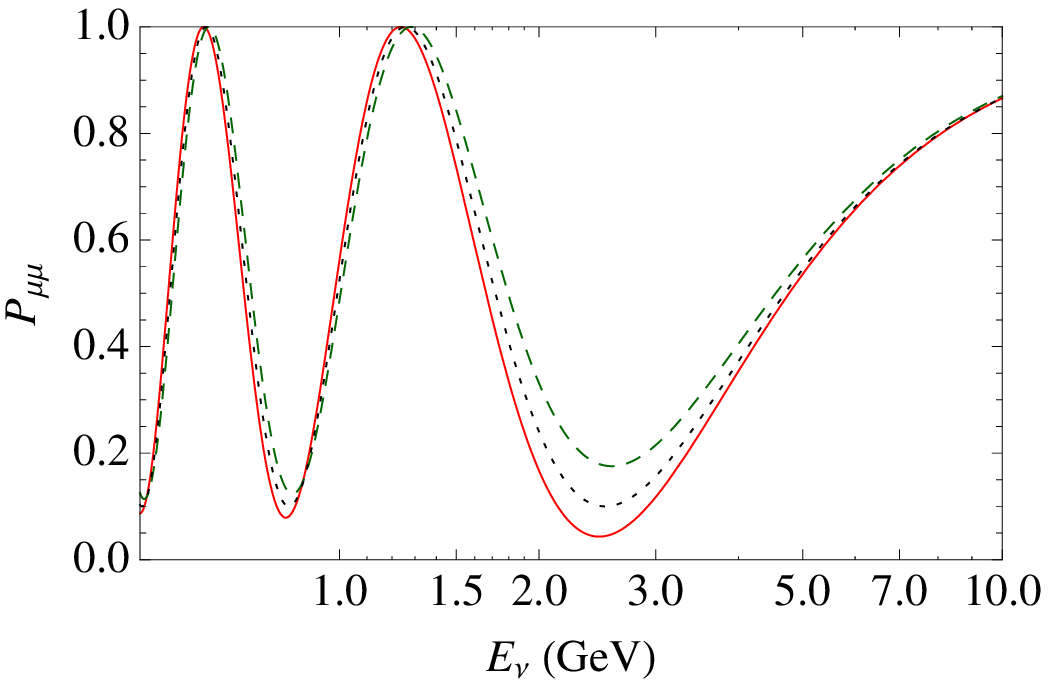} ~
\includegraphics[width=0.32\textwidth]
{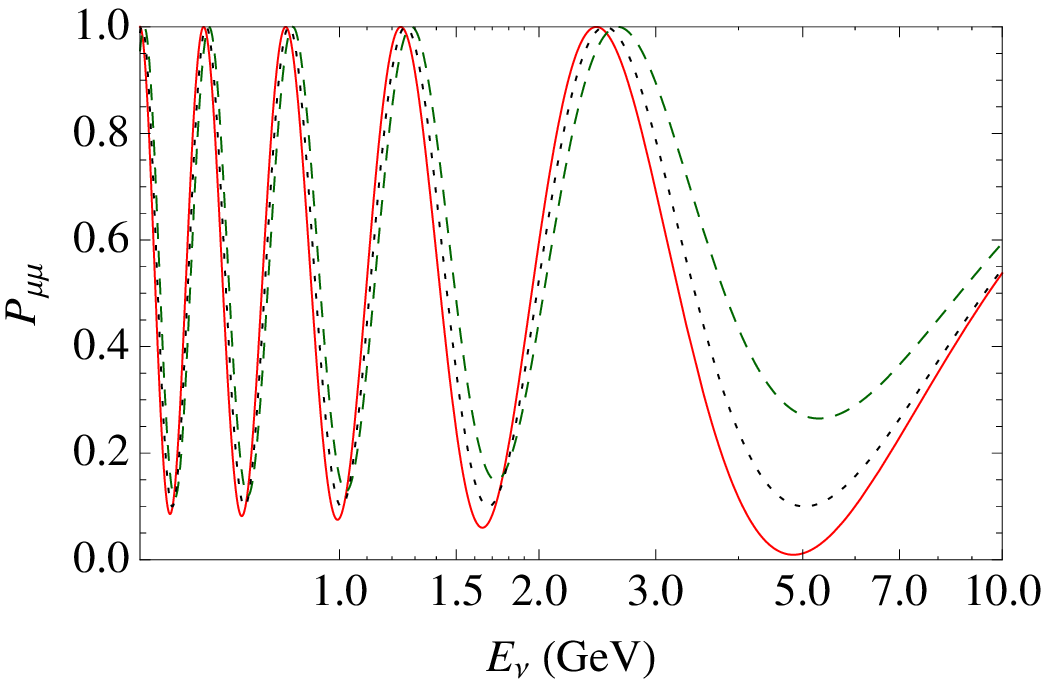} \\
(a) ~~~~~~~~~~~~~~~~~~~~~~~~~~~~~~~~~~~~~~~~~~~~~~~~~~ (b) ~~~~~~~~~~~~~~~~~~~~~~~~~~~~~~~~~~~~~~~~~~~~~~~~~~ (c)
\end{center}
\caption{
Survival probability $P_{\mu\mu}$ for $\nu_\mu$ and ${\bar\nu_\mu}$
with $E_\nu$ without (black dotted curve) and with the LRI (red solid curve for $\nu_\mu$ and green dashed
curve for ${\bar\nu_\mu}$),
for $\Delta m_{23}^2 = 2.4 \times 10^{-3}~ \text{eV}^2$,
$\sin^2 (2 \theta_{23}) = 0.9$, and $\alpha' = 1.0 \times 10^{-52}$.  Typical values for the baselines have been chosen:
(a) $L = 2 \times 6400 ~\text{km}$ (DeepCore),
(b) $L = 1300 ~\text{km}$ (DUSEL), and (c) $L = 2 \times 1300 ~\text{km}$.
The neutrino and anti-neutrino
survival probabilities are different from each other in the
presence of the LRI, since $\sin^2 (2\theta_{23}) \neq 1$.
}
\label{fig:Pmumu}
\end{figure*}

\begin{figure*}[tb]
\begin{center}
\includegraphics[width=0.32\textwidth]
{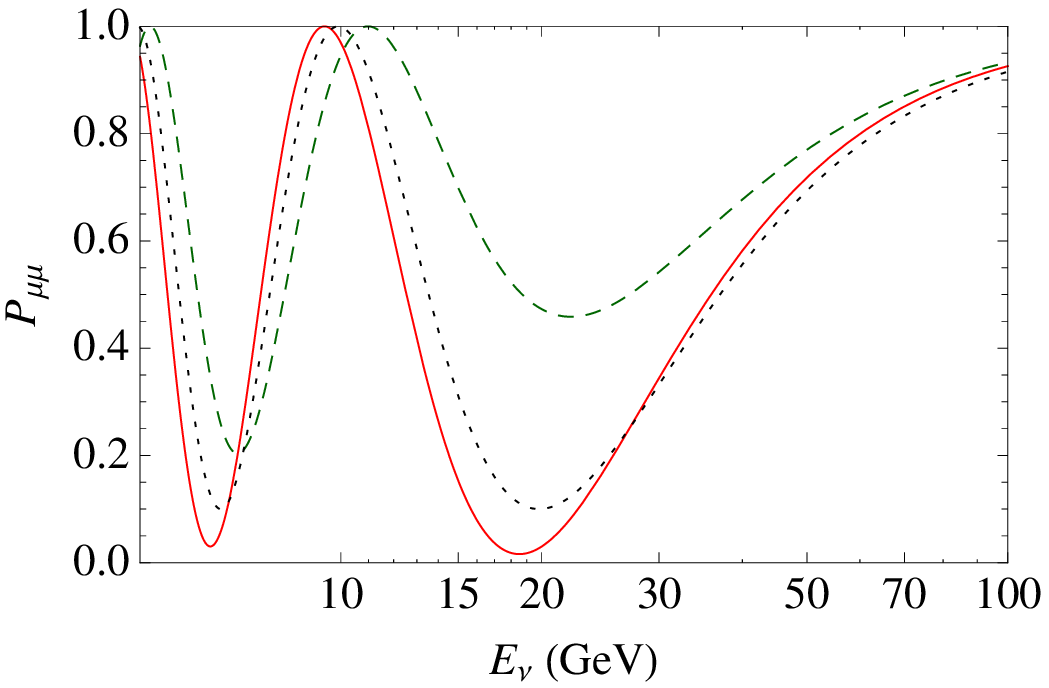} ~
\includegraphics[width=0.32\textwidth]
{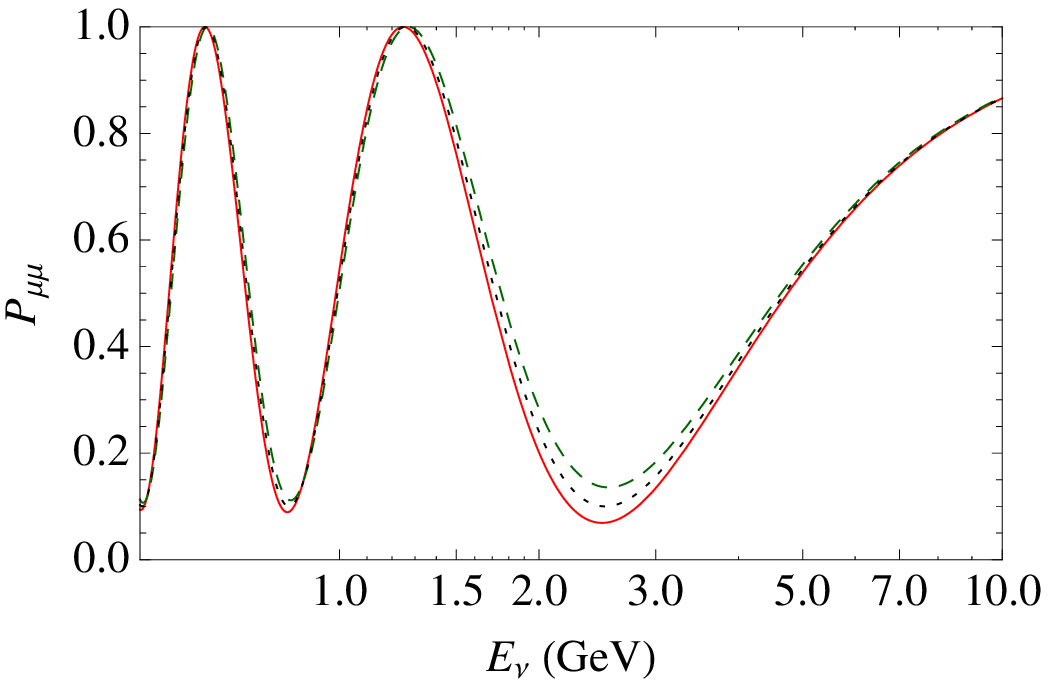} ~
\includegraphics[width=0.32\textwidth]
{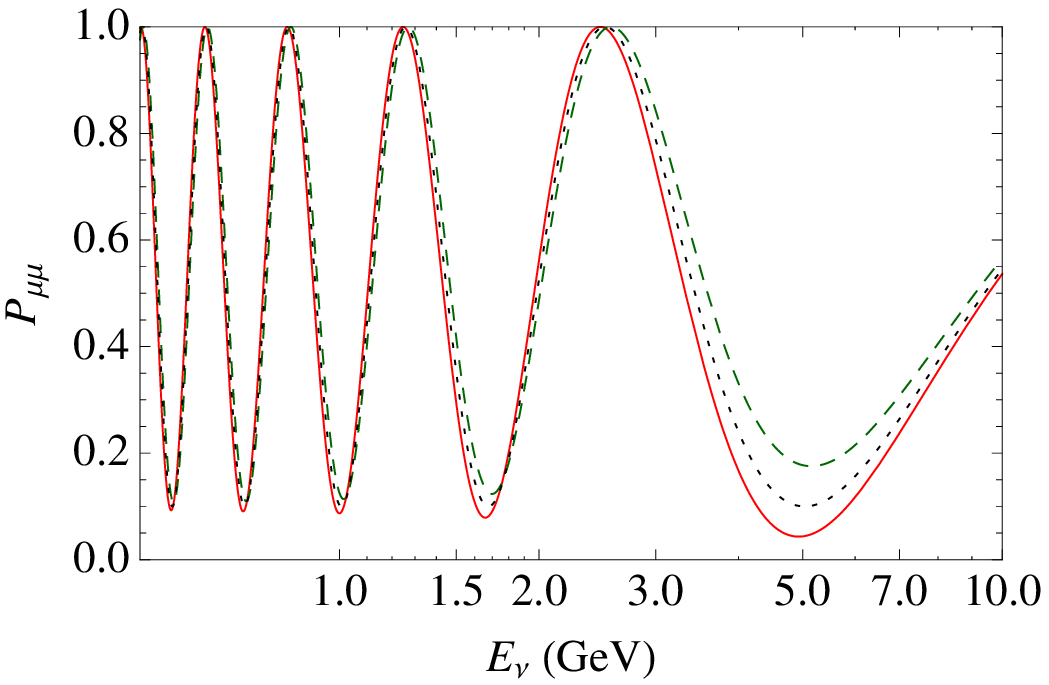} \\
(a) ~~~~~~~~~~~~~~~~~~~~~~~~~~~~~~~~~~~~~~~~~~~~~~~~~~ (b) ~~~~~~~~~~~~~~~~~~~~~~~~~~~~~~~~~~~~~~~~~~~~~~~~~~ (c)
\end{center}
\caption{
Same as Fig. \ref{fig:Pmumu} except for $\alpha' = 0.5 \times 10^{-52}$.
}
\label{fig:Pmumu2}
\end{figure*}

In this section, we will examine the implications of new LRIs for
current and future experiments.  As a guide for our following
discussion, we first derive, for the $U(1)'$ of Eq.(\ref{Q}) where neutrons are responsible 
for distinguishing $\nu_\mu$ and $\nu_\tau$, an approximate bound on $\alpha'$ based
on the MINOS $\nu_\mu$ disappearance data (which dominate the statistics \cite{MINOS}).
At 3-sigma, we roughly get $\alpha'< 5 \times 10^{-52}$, assuming $\cos(2\theta_{23}) = 0$
[corresponding to the best fit value $\sin^2(2\theta_{23})=1$
without new physics]. However, in order to address the disparity between the parameters of
neutrinos and antineutrinos suggested by the MINOS results (\ref{minosnu}) and (\ref{minosnubar}),
we will consider allowing $\cos(2\theta_{23})\neq 0$ within the LRI scenario.
We will next perform an approximate fit of the aforementioned
MINOS results, obtained at
a baseline of $L=735$~km, within our reference model.
Given the low statistical significance
of the anti-neutrino results ($\sim 100$ events \cite{MINOS}),
the fit is dominated by the neutrino data points ($\sim 2000$ events \cite{MINOS}).
For convenience, we employ the simplified MINOS data used in Ref.~\cite{Kopp:2010qt},
subtracting the neutral current background from the oscillated signals.
For the entire neutrino and anti-neutrino data set (23 bins) we find the best fit vacuum parameters:
\beq
\Delta m_{23}^2 = 2.4 \times 10^{-3}{\rm~eV}^2\; ; \;
\sin^2 (2\theta_{23}) = 0.89
\label{MINOSnuparam}
\eeq
and
\beq
\alpha' = 1.0 \times 10^{-52}\,,
\label{MINOSalpha'}
\eeq
with $\chi^2 = 20.4$ for 20 degrees of freedom.
The above fit represents the preferred values of parameters in the presence of the LRI,
although we find that the goodness of fit is basically the same
as the standard oscillations with no new physics; this was also the case
in Ref.~\cite{Heeck:2010pg}, where a  fit but with a
larger effective coupling was obtained.  However, the parameters in
Eqs.~(\ref{MINOSnuparam}) and (\ref{MINOSalpha'}) capture
the implications of a new physics effect on the $\nu_\mu$ and $\bar\nu_\mu$ data.
Next, we will examine the implications of existing bounds for our fit to the MINOS results.

\begin{figure*}[tb]
\begin{center}
\includegraphics[width=0.44\textwidth]
{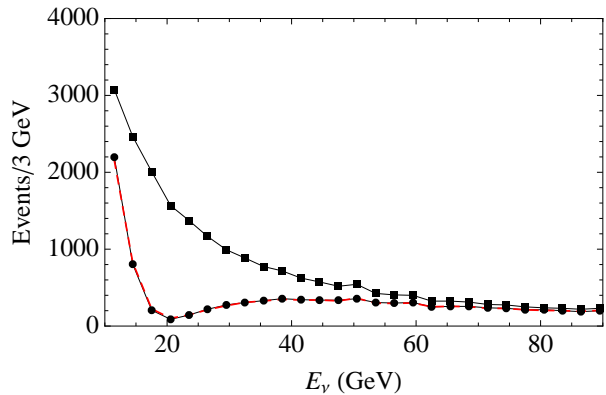}
\end{center}
\caption{
The black solid curves are numbers of upward moving atmospheric neutrino events per year at IceCube DeepCore 
as a function of energy, from Ref.~\cite{Wiebusch:2009jf}.
The upper black curve corresponds to no oscillation and 
the lower black curve accounts for standard oscillations with 
$\sin^2 (2\theta_{23}) = 1.00$ and $\Delta m_{23}^2=2.4\times 10^{-3} ~\text{eV}^2$. 
The red dashed curve is our reproduction of the standard oscillation, with the assumption of isotropic atmospheric neutrino flux with $0.68 \pi \le \phi \le \pi$, which is essentially indistinguishable from the result of Ref.~\cite{Wiebusch:2009jf}.
}
\label{fig:IceCube}
\end{figure*}

First, let us consider the constraints from the solar
and KamLAND data.  The bound obtained in Refs.~\cite{GonzalezGarcia:2006vp,Bandyopadhyay:2006uh}
by comparing KamLAND and solar neutrino data leads roughly, for our neutron (rather than electron) based
potential, to the increased bound $\alpha' < 6\times 2.5\times 10^{-53}/\cos(2\theta_{23})=
1.5\times 10^{-52}/\cos(2\theta_{23})$ at 3 sigma; the factor of 6 comes from $N_n^\odot/N_e^\odot\simeq 1/6$,
with $N_e^\odot$ the number of electrons in the Sun.
For our value of $\cos(2\theta_{23})\simeq 0.3$, the bound becomes $\alpha' < 5\times 10^{-52}$
which is about the same as the rough MINOS bound given above and 
easily satisfied by our new physics scenario.
To see how the quantity $\cos (2\theta_{23})$ enters into the bound
with our choice of gauged quantum number, note that
the solar neutrino oscillations can be described by two flavors: $\nu_e \leftrightarrow \nu_x$,
where $\nu_x \equiv \cos\theta_{23} \nu_\mu - \sin\theta_{23} \nu_\tau$, 
assuming $\theta_{13}\to 0$ [the present bound is $\sin^2(2\theta_{13}) < 0.15$, 
at 90\% confidence level \cite{Nakamura:2010zzi}].
A third eigenstate $\nu_y \equiv \sin\theta_{23} \nu_\mu + \cos\theta_{23} \nu_\tau$ decouples
in this limit (for more details see  Ref.~\cite{Bandyopadhyay:2006uh}).
With our choice of ${\cal Q} = B - L_e - 2 L_\tau$, we get
\begin{eqnarray}
\langle \nu_e | H_\text{LRI} | \nu_e \rangle &\propto& \langle \nu_e | {\cal Q}_e | \nu_e \rangle = -1 \\
\langle \nu_x | H_\text{LRI} | \nu_x \rangle &\propto& \langle \nu_x | {\cal Q}_x | \nu_x \rangle = - 2 \sin^2 \theta_{23},
\end{eqnarray}
where $H_\text{LRI}$ is the contribution of the LRI to the Hamiltonian.
Since subtracting a matrix proportional to the identity in
the evolution equation does not alter the oscillations, we have effectively
\beq
H_\text{LRI}^\text{eff} \propto \mat{0 & 0 \\ 0 & \cos 2\theta_{23}}
\eeq
for $\nu_e - \nu_x$ oscillations, which yields the aforementioned suppression by
$\cos (2\theta_{23})$.

We now turn to the atmospheric constraints, which turn out to be the tightest.  The effects of new
physics on neutrino oscillations are often discussed in terms of
coefficients $\eps_{\ell\ell'}$ \cite{Friedland:2004ah} which parametrize the strength of
the additional contributions in units of $\sqrt{2}\, G_F n_e$, the
standard MSW matter effect, with $G_F$ the Fermi constant and $n_e$ the
electron number density of the relevant medium.  The analyses in
Ref.~\cite{Friedland:2004ah} yields, at the 95\% confidence level,
the upper bound $\eps_{\tau\tau}\lsim 0.2$.
For atmospheric neutrinos traveling through the Earth, assuming an
average density of roughly 6 g cm$^{-3}$, corresponding
to $n_e \approx 1.4 \times 10^{10} ~\text{eV}^3$, that constraint implies $W_\tau = \eps_{\tau\tau}
\sqrt{2}\, G_F n_e < 4.6\times 10^{-14}$~eV.  
Our MINOS fit corresponds to $W_\tau \approx 5.6 \times
10^{-14}$~eV.  The value of $W_\tau$ from the LRI is however
approximately constant throughout the
Earth, due to the dominance of the solar contribution, whereas the density of
the Earth, sampled by neutrinos traveling along its diameter,
varies from roughly 12 g cm$^{-3}$, corresponding to $W_\tau \sim 9 \times 10^{-14}$~eV,  
in the core (which extends to $\sim 3400$~km from the center of the Earth) 
to 5 g cm$^{-3}$, corresponding to $W_\tau \sim 4 \times 10^{-14}$~eV, in the
mantle (whose thickness is $\sim 2900$~km).    
Thus, the mean potential sampled by neutrinos traveling close to 
the diameter will be somewhere in between these values, 
roughly $W_\tau \sim 6 \times 10^{-14}$~eV, somewhat larger than 
the value corresponding to our MINOS fit.    
Therefore, while our fit appears to have some tension
with the atmospheric constraint, a more detailed analysis is called for, in order to
pinpoint a precise constraint on the contribution from the LRI we have assumed.
Henceforth, we will adopt $\alpha' = 1.0 \times 10^{-52}$ as
a plausible benchmark value for further exploration by
experiments, where the potential MINOS anomaly suggests that new physics may appear.
For our numerical illustrations, we will use vacuum parameters
$\Delta m_{23}^2 = 2.4 \times 10^{-3}{\rm~eV}^2$ and
$\sin^2(2\theta_{23}) = 0.9$, which are motivated
by our fit results in Eq.~(\ref{MINOSnuparam}) and are also
reasonable in the standard scenario without new physics.  In this way, we can focus on the effect of
the LRI potential on neutrino oscillations, given specific vacuum parameters that
could in principle be determined with high precision from other measurements.

\begin{figure*}[tb]
\begin{center}
\includegraphics[width=0.44\textwidth]
{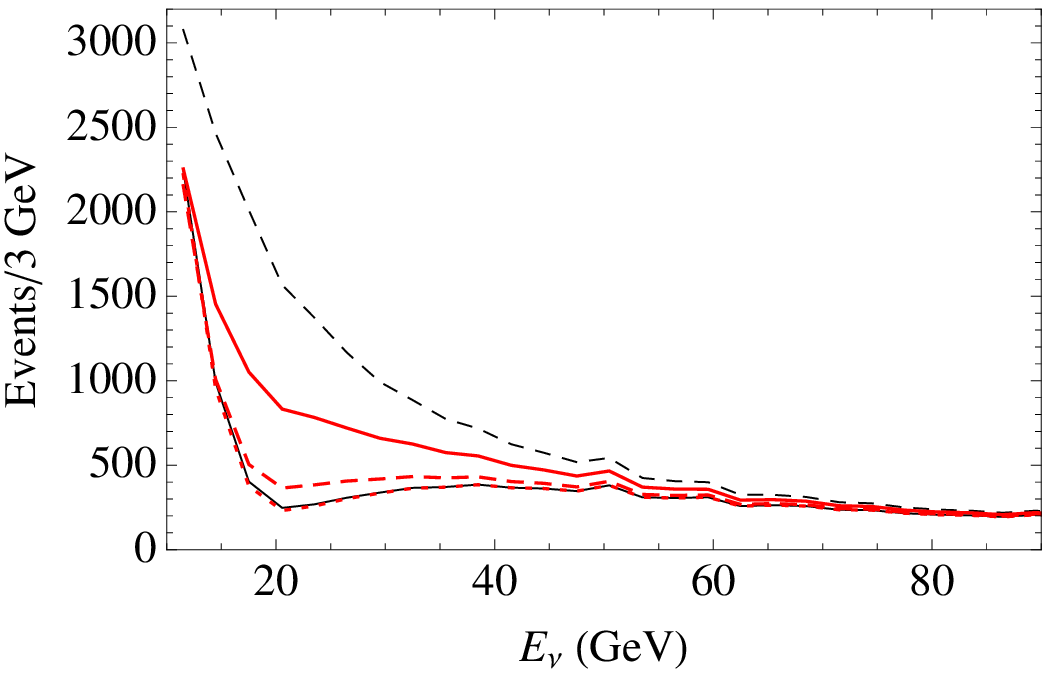} ~~
\includegraphics[width=0.44\textwidth]
{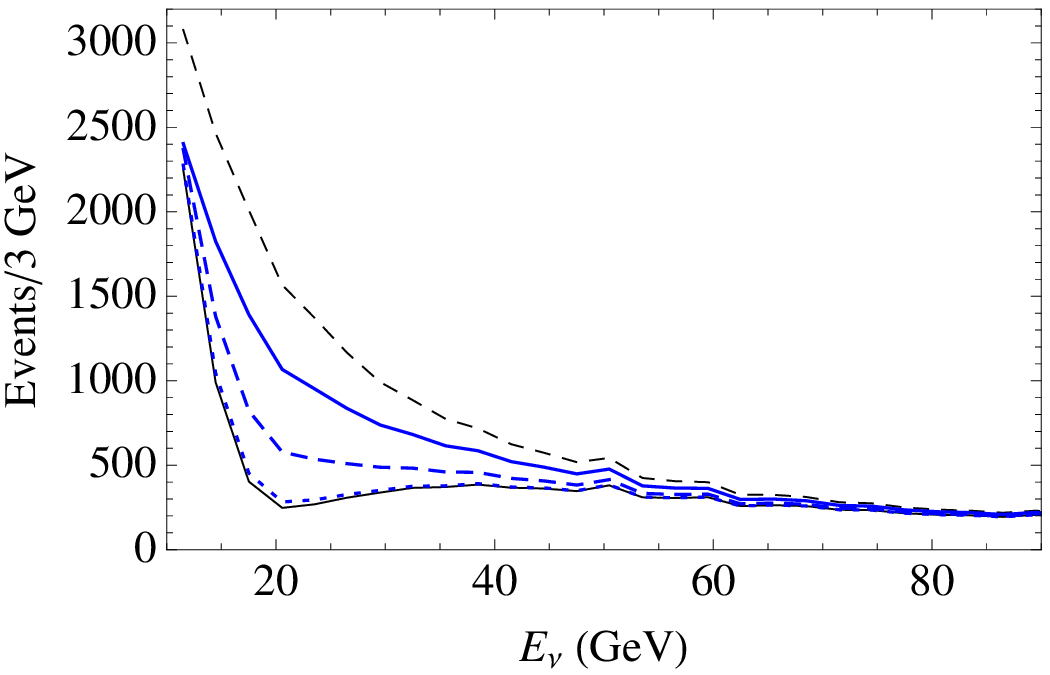}  \\
(a) ~~~~~~~~~~~~~~~~~~~~~~~~~~~~~~~~~~~~~~~~~~~~~~~~~~~~~~~~~~~~~~~~~~~~~~ (b)
\end{center}
\caption{Atmospheric neutrinos per year (per 3 GeV) with $E_\nu$ at the IceCube DeepCore, 
using the same assumptions applied for Fig.~\ref{fig:IceCube}.
Shown are the unoscillated case (top black dashed curves) and the case of no new physics (bottom thin black solid curves), as well as the cases $\alpha' = 1.0, 0.5, 0.1 \times 10^{-52}$
corresponding to thick solid,
dashed, and dotted curves, respectively.  The vacuum parameters are
$\sin^2 (2\theta_{23}) = 0.9$ with (a) $\Delta m_{23}^2=2.4\times 10^{-3} ~\text{eV}^2$ and (b) $\Delta m_{23}^2=-2.4\times 10^{-3} ~\text{eV}^2$.
}
\label{fig:DEEPCORE}
\end{figure*}

We also mention that the same type of study outlined above can easily be applied to other 
$U(1)'$ gauge groups with different quantum numbers and LRIs.  Two interesting examples, gauged 
$L-3L_\tau$ and $B-3L_\tau$, turn out to have essentially the same effect on MINOS as well as 
atmospheric and long baseline neutrinos as our $B-L_e-2L_\tau$ example, but with smaller values for  
$\alpha'$ of $1.3\times 10^{-53}$ and $1.1\times 10^{-53}$, respectively.  Here, the smaller values of $\alpha'$ 
are compensated by the larger numbers of electrons and baryons.

With the completion of the DeepCore array, the IceCube experiment will be able to
probe atmospheric neutrino oscillations for energies of order 10~GeV and above,
well beyond the typical reach of the SuperKamiokande detector.  The expected large
statistics, of order $10^5$ events per year \cite{Wiebusch:2009jf}, make DeepCore an interesting
probe of LRI using atmospheric neutrino data, which we now consider.

Fig. \ref{fig:Pmumu}(a) shows the $\nu_\mu$ survival probability
$P_{\mu\mu}$ versus energy $E_\nu$ without (black dotted curve) and
with the LRI, where the red solid and green dashed curves
correspond to $\nu_\mu$ and $\bar\nu_\mu$, respectively. We have
assumed the Earth diameter as the baseline, $L = 2 \times 6400
~\text{km}$, as a typical value relevant to the DeepCore array.
$\Delta m_{23}^2 = 2.4 \times 10^{-3}~ \text{eV}^2$, $\sin^2 (2
\theta_{23}) = 0.9$, and $\alpha' = 1.0 \times 10^{-52}$ have been
chosen for illustration purposes.  The sign of $\Delta m_{23}^2$
has been chosen according to our fit in Eq.~(\ref{MINOSnuparam}),
and for definiteness we will choose $\cos(2\theta_{23}) >0$ throughout our analysis.
We see that the LRI distinguishes
neutrinos and anti-neutrinos for $\sin^2 (2 \theta_{23}) \neq 1$.
The deviation of the LRI from the standard scenario
becomes significant for $E_\nu \gsim 15~\text{GeV}$ for neutrinos, whereas
the effect is quite large for anti-neutrinos over  the entire
range of energies considered here.

Fig. \ref{fig:Pmumu}(b) shows a similar plot
for $L = 1300 ~\text{km}$, typical of the baseline for DUSEL
experiments, which happens to be approximately $1/10$ of the Earth diameter, with the other parameters as in panel (a).
Deviations from standard oscillations do not become as significant as in case
(a) for the same value of $E_\nu/L$, due to the additional energy
dependence of the LRI parameters in Eqs.~(\ref{m23til}) and
(\ref{sin23til}).  The difference between the $\nu_\mu$ and
$\bar \nu_\mu$ signal is larger than their individual deviation from
standard oscillations and can be a potentially distinct signal.

Fig. \ref{fig:Pmumu}(c) shows the effect of increasing the baseline by a factor of two,
$L = 2 \times 1300 ~\text{km}$, compared to the case presented in Fig. \ref{fig:Pmumu}(b).
For the given energy range, as expected from Eq.~(\ref{Pmumu}), more oscillations will take place.
We also see that the effect of the LRI on oscillations is much more prominent, but at about
twice the value of $E_\nu$ compared to that in Fig. \ref{fig:Pmumu}(b).  It should
be noted that merely increasing the baseline by a factor of two compared to the case in Fig. \ref{fig:Pmumu}(b)
would result in a reduction by $1/4$ in the flux and hence the event rate,
all other factors being equal.

The results in
Fig. \ref{fig:Pmumu} suggest that it might be easier to see the LRI
effect in the DeepCore experiment than at a long-baseline experiment.
On the other hand, long-baseline experiments, unlike the
DeepCore, can in principle detect the asymmetry in the  $\nu_\mu$
and $\bar \nu_\mu$ oscillations for $\sin^2 (2\theta_{23}) \ne
1$, which is a key feature of the LRI assumed here.

Fig. \ref{fig:Pmumu2} shows the same qualitative features for a
smaller coupling $\alpha' = 0.5 \times 10^{-52}$; as expected, the
effect is less pronounced. In any event, the plots suggest that for
$E_\nu \gsim 15$~GeV DeepCore can be quite sensitive to the new
physics.  Note that the effect on $\nu_\mu$ and $\bar\nu_\mu$
can be interchanged by changing the sign of $\Delta m_{23}^2$.

To estimate the reach for LRI 
effects in neutrino oscillations at the DeepCore array, we will only consider the upward 
moving flux of atmospheric 
neutrinos.  We use the no-oscillation  
number of events per year   
as a function of energy, 
provided by the IceCube collaboration in Ref.~\cite{Wiebusch:2009jf} (where these 
events are referred to as ``vertical'').  This event rate is represented by  
the top black solid curve (adapted from 
Ref.~\cite{Wiebusch:2009jf}) 
in Fig. \ref{fig:IceCube}.  The lower black solid curve in this figure is the  
expected number of upward moving events for standard oscillations with 
$\sin^2 (2\theta_{23}) = 1.00$ and $\Delta m_{23}^2=2.4\times 10^{-3} ~\text{eV}^2$, 
also provided by the IceCube collaboration \cite{Wiebusch:2009jf}.  In order to reproduce 
this curve, we assume an isotropic flux of atmospheric neutrinos at each point around the 
Earth, and vary the zenith angle interval $\phi_0 \leq \phi \leq \pi$  
over which we calculate the event rate for upward moving neutrinos with oscillations, 
for the same values of $\sin^2 (2\theta_{23})$ 
and $\Delta m_{23}^2$ as above.
The zenith angle $\phi = \pi$ ($\pi/2$) corresponds to the neutrinos moving vertically upward (horizontally) towards the detector.
We have also assumed the ratio of $\nu_\mu$ to $\bar \nu_\mu$ events with no oscillations to be $2:1$, reflecting a good approximation for the ratio of the cross sections \cite{Zeller:2001hh}.     
The red dashed curve in Fig.~\ref{fig:IceCube} is our reproduction of the oscillation with $\phi_0=0.68 \pi$, which agrees very well with the solid curve (from Ref.~\cite{Wiebusch:2009jf}) and is basically indistinguishable from it in the plot.
Hence, we will use the above approximations, with $\phi_0=0.68\pi$, to estimate the reach for new LRI effects at the DeepCore array, in the following.

\begin{figure*}[tb]
\begin{center}
\includegraphics[width=0.44\textwidth]
{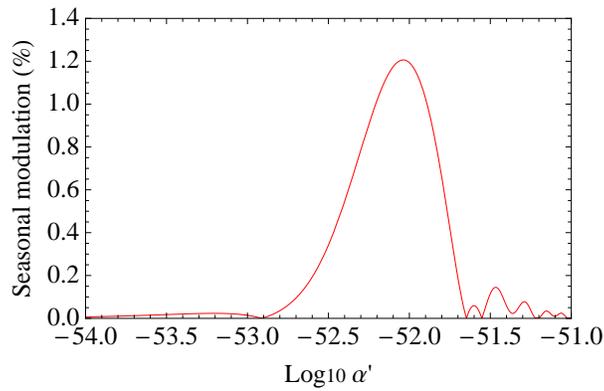}
\end{center}
\caption{
Percentage of annual modulation $|(N_a - N_p) / (N_a + N_p)|$ in atmospheric neutrino oscillation at DeepCore versus the LRI coupling $\alpha'$, for $15 < E_\nu ~(\text{GeV}) < 30$ around aphelion and perihelion.  For each season, 120 days have been included. Other parameter values are the same as those of Fig.~\ref{fig:DEEPCORE}(a).
}
\label{fig:DEEPCOREannual}
\end{figure*}

Fig. \ref{fig:DEEPCORE}(a) illustrates the effect of the LRI on the
number of events (for a one-year run and per 3 GeV energy bins) in
$\nu_\mu$ and $\bar \nu_\mu$ disappearance experiments.  The top (dashed) curve
corresponds to no oscillations \cite{Wiebusch:2009jf}.  The thick (red) solid, dashed, and dotted curves
are for $\alpha' = 1.0, 0.5, 0.1 \times 10^{-52}$, respectively and
$\alpha'=0$ (no new physics) is represented by the thin (black) solid curve at the bottom.  
The vacuum parameters
are $\sin^2 (2\theta_{23}) = 0.9$ and $\Delta m_{23}^2 = 2.4 \times 10^{-3}$~eV$^2$,
for all cases, to illustrate the effect of the LRI on neutrino oscillations.  We see that except for the
smallest value of $\alpha'$ the other cases are very distinct from the no new physics case,
for 15 $\lsim E_\nu~(\text{GeV}) \lsim$ 40.  Over this range of energies, the no-new-physics case
yields roughly 3680 events and for $\alpha' = 0.1 \times 10^{-52}$
the number of events changes by about 100.  Hence, statistically, a few-percent-level measurement,
which might require 3 years of data, could in principle
reach one order of magnitude below our benchmark value of $\alpha' = 1.0 \times 10^{-52}$,
at the 3 sigma level.

Fig. \ref{fig:DEEPCORE}(b) is the same as Fig. \ref{fig:DEEPCORE}(a) except that
$\Delta m_{23}^2 = - 2.4 \times 10^{-3}$~eV$^2$, interchanging $\nu$ and $\bar\nu$,
is chosen.  The thick (blue)
solid, dashed, and dotted lines correspond to the same values of $\alpha'$ as in
Fig. \ref{fig:DEEPCORE}(a)  and the thin black solid curve at the bottom
represents no new physics.
We see that the effect of the LRI is now distinct for all values of $\alpha'$.
For 15 $\lsim E_\nu~(\text{GeV}) \lsim$ 40,  the number of events for the new-physics case
with $\alpha' = 0.1 \times 10^{-52}$ differs from that of the no-new-physics case by about 220.
We see that even a several-percent-level measurement could in principle
reach one order of magnitude below our benchmark value of $\alpha' = 1.0 \times 10^{-52}$.  Given the
large statistical samples expected at DeepCore, our estimates suggest that
values of $\alpha'$ around one order of magnitude below that
of our benchmark $\alpha'$ could potentially be probed by this experiment.

\begin{figure*}[tb]
\begin{center}
\includegraphics[width=0.44\textwidth]
{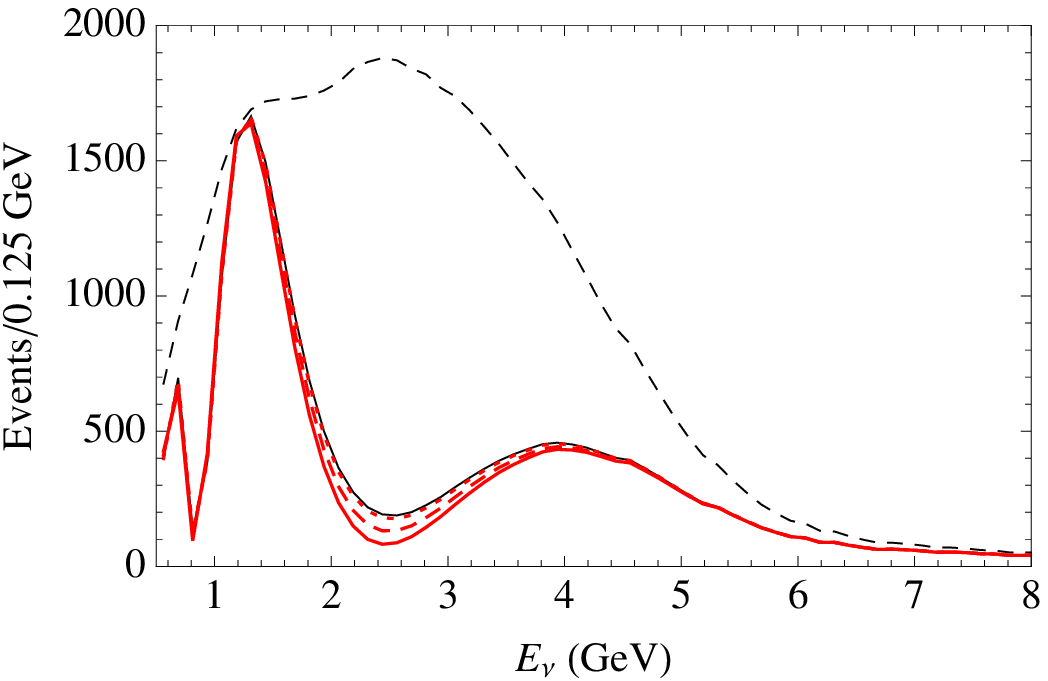} ~~
\includegraphics[width=0.44\textwidth]
{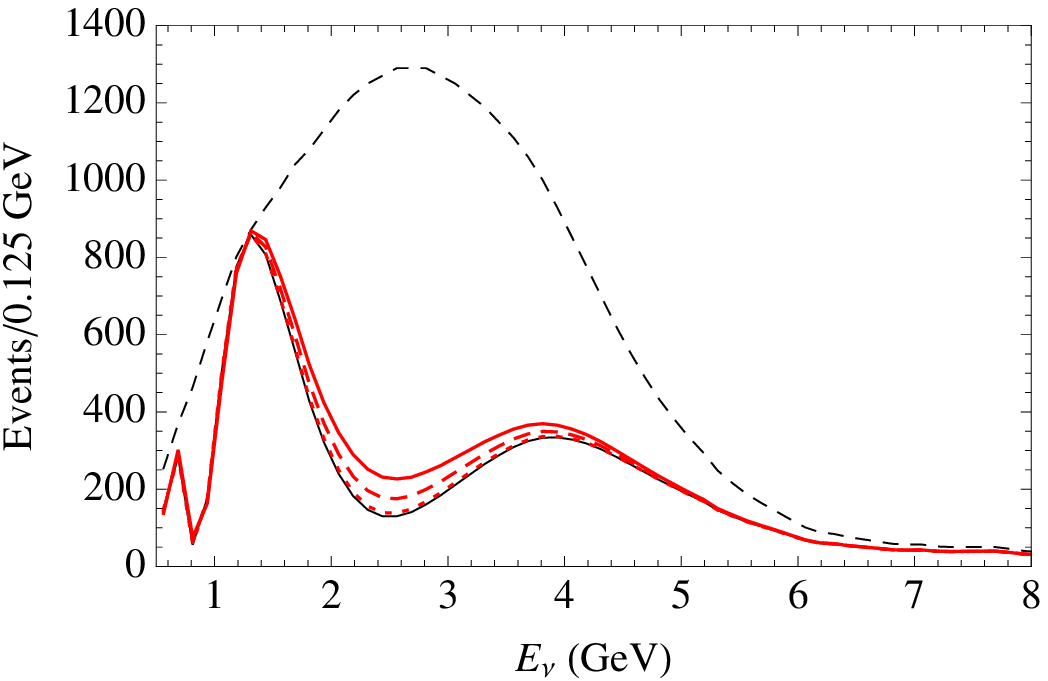} \\
(a) ~~~~~~~~~~~~~~~~~~~~~~~~~~~~~~~~~~~~~~~~~~~~~~~~~~~~~~~~~~~~~~~~~~~~~~ (b)
\end{center}
\caption{The number of (a) neutrino and (b) anti-neutrino events for a 5-year run
(per 0.125 GeV) versus $E_\nu$, in a long-baseline experiment with $L = 1300$ km (DUSEL).
The unoscillated case (top black dashed curves) and the case of no new physics (thin black solid curves)
are displayed, as well as the cases with $\alpha' = 1.0, 0.5, 0.1 \times 10^{-52}$ corresponding to thick solid,
dashed, and dotted curves, respectively.  The vacuum parameters are
$\sin^2 (2\theta_{23}) = 0.9$ and $\Delta m_{23}^2=2.4\times 10^{-3} ~\text{eV}^2$.
}
\label{fig:DUSEL}
\end{figure*}

\begin{figure*}[tb]
\begin{center}
\includegraphics[width=0.44\textwidth]
{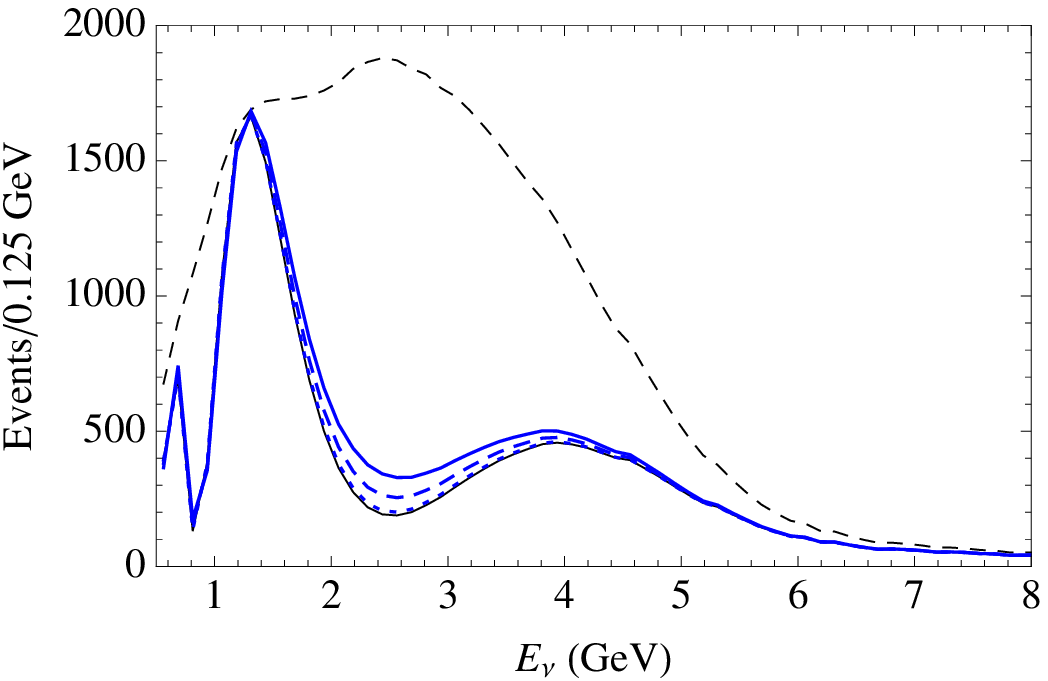} ~~
\includegraphics[width=0.44\textwidth]
{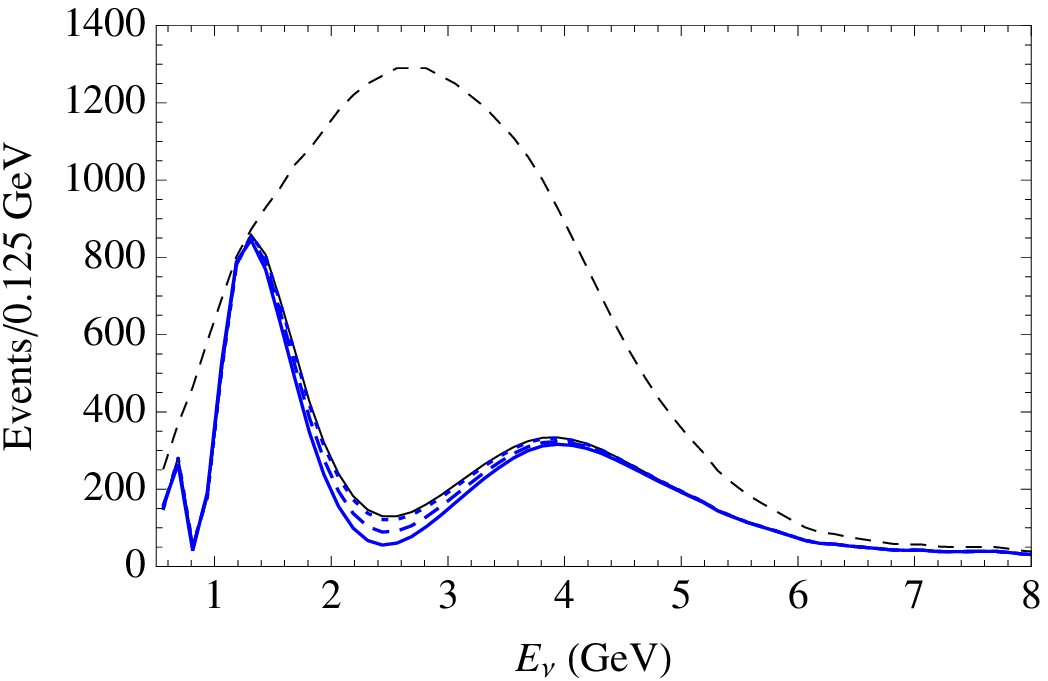} \\
(a) ~~~~~~~~~~~~~~~~~~~~~~~~~~~~~~~~~~~~~~~~~~~~~~~~~~~~~~~~~~~~~~~~~~~~~~ (b)
\end{center}
\caption{Same as Fig. \ref{fig:DUSEL} except for $\Delta m_{23}^2=-2.4\times 10^{-3} ~\text{eV}^2$.
}
\label{fig:DUSELII}
\end{figure*}

Fig. \ref{fig:DEEPCOREannual} shows our estimate for the size of the annual
modulation of atmospheric neutrino oscillations at DeepCore, as a function
of the LRI coupling $\alpha'$.  Here, the vertical
axis is $|(N_a - N_p) / (N_a + N_p)|$, where $N_a$ and $N_p$ are
the numbers of events associated with aphelion and perihelion, respectively.
An energy cut of $15 < E_\nu ~(\text{GeV}) < 30$ has been implemented and
120 days have been included around each apsis.  For our estimates, we have simulated
the variation in $\res$ by a sinusoidal function.  This approximation of
Earth's true Keplerian orbit captures the main effect we would like to illustrate, at
the level of our analysis.   The same parameter values as in
Fig.~\ref{fig:DEEPCORE}(a), {\it i.e.} $\Delta m_{23}^2 = 2.4 \times 10^{-3}~
\text{eV}^2$ and $\sin^2(2 \theta_{23}) = 0.9$ are assumed.
The total number of events $(N_a + N_p)$ per year for $\alpha' = 1.0 \times 10^{-52}$ and
$\alpha' = 0.5 \times 10^{-52}$ are about 2700 and 1400, respectively,
and the seasonal modulations are near $1.2 \%$ for the former and $0.8 \%$ for the latter.
Our estimates then suggest that, depending on the value of $\alpha'$, 3-10 years of data
could yield the necessary statistics to measure such levels of modulation.
We have not accounted for atmospheric neutrino flux uncertainties,
which can be as much as $10 - 15\%$.  However, the
$(\nu_\mu + {\bar \nu_\mu})$/$(\nu_e + {\bar \nu_e})$ flux ratio, which is
proportional to our result, can be known much more precisely and
would have uncertainties at the $1 - 2\%$ level \cite{Barr:2006it}.  The large amount of
statistics expected at DeepCore makes per-cent level measurements a realistic
possibility \cite{FernandezMartinez:2010am}.  We may also expect that a more detailed and optimized
analysis of the real data, using the predicted time evolution of a stable flux ratio,
could allow for a larger observed effect.  Hence, our estimate suggests that DeepCore
could be sensitive to annual modulations, when the solar source particles dominate the
LRI potential, as we have assumed in this work.

\begin{figure*}[tb]
\begin{center}
\includegraphics[width=0.44\textwidth]
{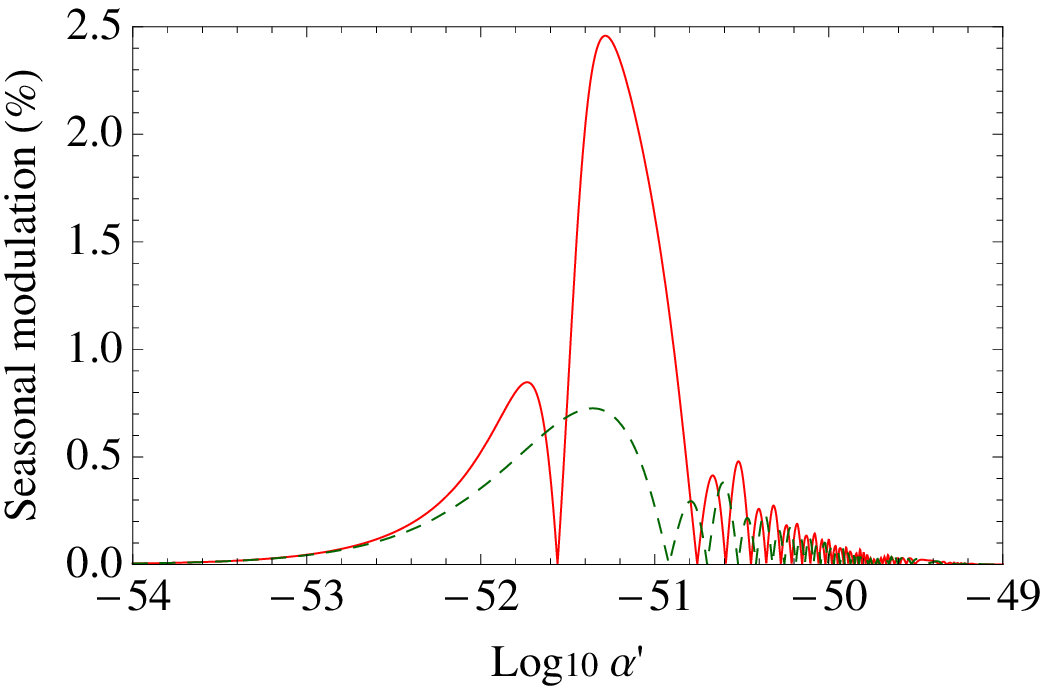}
\end{center}
\caption{Percentage of annual modulation $|(N_a - N_p) / (N_a + N_p)|$
for a 5-year run versus the LRI coupling $\alpha'$, for $2 < E_\nu ~(\text{GeV}) < 3$
around aphelion and perihelion.  In each season, 180 days have been
included, for $\nu_\mu$ (red solid curve) and $\bar \nu_\mu$ (green dashed curve).
The same parameter values as Fig. \ref{fig:DUSEL} are used.
}
\label{fig:DUSELseasonal}
\end{figure*}

Fig. \ref{fig:DUSEL} shows the predictions for the number of
events per 0.125 GeV energy bins, over a 5-year run, for (a)
$\nu_\mu$ and (b) $\bar \nu_\mu$ long-baseline
disappearance experiments, {\it e.g.} at DUSEL,
with $L = 1300$ km. We have taken the unoscillated beam profile
(dashed black curves at the top) from Ref.~\cite{DUSEL}, corresponding to a
200~kt water \v{C}erenkov detector.  The vacuum parameters are $\sin^2(2\theta_{23}) =0.9$
and $\Delta m_{23}^2 = 2.4 \times 10^{-3}~\text{eV}^2$.
The thin black solid curves correspond to no new physics.  The thick red solid, dashed, and dotted
curves again correspond to $\alpha'=1.0, 0.5, 0.1 \times 10^{-52}$.
We see that the largest two values of $\alpha'$ yield predictions that are distinct
from no new physics, for 2 $\lsim E_\nu ~(\text{GeV}) \lsim$ 3 in both
neutrino and anti-neutrino cases.  We see that the LRI leads to distinct effects for $\nu_\mu$ and
$\bar \nu_\mu$, which is a key signature of this new physics.  However,
the smallest value of $\alpha'$ does not yield a visible effect on these plots and may be difficult to
reach at these experiments.

Fig. \ref{fig:DUSELII} contains the same information, except for
$\Delta m_{23}^2 = - 2.4 \times 10^{-3}~\text{eV}^2$,
with the same conventions (new physics contributions are now displayed with blue thick lines).  The same
qualitative features as in the previous case with $\Delta m_{23}^2 >0$ are present and reaching
$\alpha'\sim 0.1 \times 10^{-52}$ seems to be difficult here as well.

Fig. \ref{fig:DUSELseasonal} shows the seasonal modulation for a
5-year run of the DUSEL as a function of the LRI coupling $\alpha'$ for
$\nu_\mu$ (red solid curve) and $\bar \nu_\mu$ (green dashed curve). We take $2 < E_\nu
~(\text{GeV}) < 3$, with 180 days around each apsis.
The same parameter values as in Fig. \ref{fig:DUSEL},
$\sin^2(2 \theta_{23}) = 0.9$, $\Delta m_{23}^2 = 2.4 \times
10^{-3}~ \text{eV}^2$, are assumed.
For $\alpha' = 1.0 \times 10^{-52}$, the total number of events $(N_a + N_p)$ are about 1100 ($\nu_\mu$)
and $2100$ ($\bar \nu_\mu$)
with the modulation at $0.5 \%$ and $0.4 \%$, respectively.
For $\alpha' = 0.5 \times 10^{-52}$, the modulation is at $0.2\%$, for
the total number of events about 1500 ($\nu_\mu$) and $1700$ ($\bar \nu_\mu$), respectively.
We assume the beam flux is constant over time.  While our analysis
is only a rough estimate, we may conclude that observing the modulation at  the per-cent level
would require experiments (beams and detectors) with somewhat more enhanced capabilities,
compared to those assumed for this analysis.

\section{Summary and Conclusions}
\label{conclusions}
In this work, we examined the effect of long-range interactions on neutrino oscillation
experiments, motivated by the recent accelerator data from MINOS.  These data are
not conclusive, but could be suggesting that
the oscillation parameters for neutrinos and anti-neutrinos may be distinct.  Such an effect, if
confirmed with more data in the future, could in principle be caused by long range
interactions coupled to neutrinos \cite{JM,Grifols:2003gy,Heeck:2010pg}.
As an illustrative example, we considered a $U(1)'$ model with an ultra light gauge boson
coupled to $(B-L )+ (L_\mu-L_\tau)=B-L_e-2L_\tau$ 
that captures the key aspects of the requisite
phenomenology.  Such an effective interaction can also arise in other ways,
for example through gauge boson mixing  from two separate sectors \cite{Heeck:2010pg}.
The main required features are an interaction with a range of order 1~AU
and coupling to both stellar matter and neutrinos, with the resulting potential characterized
by a fine structure constant $\alpha'\leq 10^{-52}$.
Alternatively, for a shorter range interaction, $m_{Z'} \gsim 10^{-16}$ eV, the Earth itself may be the dominant source of terrestrial neutrino potential differences.
We pointed out that when the Sun
is the dominant  source of the long-range potential, the effect on neutrino oscillations
at the Earth will exhibit annual modulations, due to the variable distance
between the Earth and the Sun.

We performed an approximate fit to the MINOS data within our reference model
that gave a qualitative description of the data.
Our fit results accommodate the current experimental bounds on
non-standard contributions to neutrino oscillation data, although they show some tension with atmospheric bounds.
However, for benchmark parameters motivated by our MINOS fit, we show that ongoing and future experiments
could detect large effects due to the LRI potential,
or else significantly further constrain such new physics.  In particular, we estimated that the
currently operational IceCube DeepCore array can reach well beyond,
by about an order of magnitude, the
benchmark parameter space suggested by our MINOS fit,
with about one year of atmospheric neutrino
data at a typical baseline given by the size of the Earth, for
$15 \lsim E_\nu ~(\text{GeV}) \lsim 40$.  In addition, the large statistics afforded by
DeepCore seem sufficient to detect a per-cent level
modulation of neutrino oscillations with 3-10 years of data,
providing key evidence for the solar source of the long-range potential.

While DeepCore is only sensitive to the sum of neutrino and anti-neutrino events, they can be
separately probed at future precision long-baseline experiments,
such as those at a future DUSEL facility.  With typical assumptions about
the capabilities of such experiments, we showed that the values of
parameters motivated by the MINOS results will be well covered, with about 5 years of data.  These
experiments can in principle uncover an asymmetry in the properties of neutrinos and anti-neutrinos,
which is a key feature of the long-range interactions we have considered.  Our simple estimates
suggest the annual modulation of the data may not be easily accessible in these experiments
unless their capabilities are somewhat more enhanced compared to our reference values.  

For $U(1)'$s with gauged $L-3L_\tau$ and $B-3L_\tau$, we found essentially the same 
phenomenology but with couplings $\alpha'$ about an order of magnitude smaller because of the 
larger sources of electrons and protons in the Sun, compared to neutrons.  In all cases, DeepCore and 
future long baseline neutrino oscillation experiments are expected to push 
the $\alpha'$ sensitivity about an order 
of magnitude beyond our benchmark values.    

Before closing, we would like to comment on some related possibilities that could be
of interest in the context of the new physics considered here.  First of all, in our discussion
of annual modulation, we mainly considered future data.  However, the
Super-Kamiokande atmospheric data could already
contain the annual modulation signal in the ratio of the muon and electron
neutrinos and a dedicated analysis could be of interest in view of
the MINOS anomaly.  Also, while we largely focused on a particular type of long-range interaction, 
interesting motivation can be found in other possibilities.  For example, 
the $B - x_\ell L_\ell$ type of model, where the lepton sector
has flavor-dependent couplings, can contain residual discrete symmetries that
stabilize the proton and a dark matter candidate \cite{Lee:2010hf}.
This could provide possible new connections between neutrino experiments and
other areas of particle physics. It is conceivable that dark
matter is charged under a new long-range force \cite{Friedman:1991dj}, via a generalization of the
setup in Ref.~\cite{Lee:2008pc}, or as in Ref.~\cite{Kaloper:2009nc}.  Then, new effects
could perhaps arise from dark matter trapped in the Sun or the Earth,
in such scenarios.  (While the modulation observed at the DAMA/LIBRA experiment
\cite{Bernabei:2010mq} might be explained by the motion of the Earth through
the dark matter halo of the Galaxy, we find the relative proximity of the dates of the apses
to the extrema observed by this experiment intriguing.  A
long-range potential from solar sources that also acts on dark matter could in principle
affect the interpretation and implications of these measurements.)

Hence, we conclude that current and future atmospheric and long-baseline neutrino oscillation
experiments will provide an opportunity to probe long-range interactions whose
feebleness generally excludes other experimental search avenues.  Together,
such experiments, like those at the DeepCore array and DUSEL, can test
the relevance of these interactions to the suggested MINOS anomaly,
or place more stringent bounds on their parameters.  Detection
of a new long-range force would be an important discovery with
an immense impact on our view of fundamental physics.

\vspace{0.5cm}
\acknowledgments

We thank Milind Diwan, Alexander Friedland, Patrick Huber, Cecilia Lunardini,
and Jonghee Yoo for conversations.  This work was supported in part by the United States Department
of Energy under Grant Contracts DE-AC02-98CH10886.


\begin{thebibliography}{99}

\bibitem{GonzalezGarcia:2007ib}
  M.~C.~Gonzalez-Garcia, M.~Maltoni,
  Phys.\ Rept.\  {\bf 460}, 1-129 (2008).
  [arXiv:0704.1800 [hep-ph]].

\bibitem{Nakamura:2010zzi}
  K.~Nakamura {\it et al.}  [Particle Data Group],
  J.\ Phys.\ G {\bf 37}, 075021 (2010).

\bibitem{Wolfenstein:1977ue}
  L.~Wolfenstein,
  Phys.\ Rev.\  D {\bf 17}, 2369 (1978).

\bibitem{Mikheev:1986gs}
  S.~P.~Mikheev and A.~Y.~Smirnov,
  Sov.\ J.\ Nucl.\ Phys.\  {\bf 42}, 913 (1985)
  [Yad.\ Fiz.\  {\bf 42}, 1441 (1985)].

\bibitem{Botella:1986wy}
  F.~J.~Botella, C.~S.~Lim and W.~J.~Marciano,
  Phys.\ Rev.\  D {\bf 35}, 896 (1987).
This paper computes the SM loop-induced index of refraction
difference between $\nu_\mu$ and $\nu_\tau$ and concludes that it is about $5\times 10^{-5}$
smaller than the MSW effect in terrestrial neutrino oscillation experiments, {\it i.e.}
unobservably small, leaving room for the discovery of a new physics potential.

\bibitem{MINOS}
Talk by P. Vahle at
24th International Conference On Neutrino Physics
And Astrophysics (Neutrino 2010), June 14, 2010.
({\tt http://indico.cern.ch/getFile.py/\\access?contribId=201\&sessionId=1\&resId=0\&\\materialId=slides\&confId=73981}).
Essentially the same results have since been reported in
  P.~Adamson {\it et al.}  [MINOS collaboration],
  arXiv:1104.0344 [hep-ex].

\bibitem{Heeck:2010pg}
  J.~Heeck and W.~Rodejohann,
  J.\ Phys.\ G {\bf 38}, 085005 (2011)
  [arXiv:1007.2655 [hep-ph]].

\bibitem{Engelhardt:2010dx}
  N.~Engelhardt, A.~E.~Nelson and J.~R.~Walsh,
  Phys.\ Rev.\  D {\bf 81}, 113001 (2010)
  [arXiv:1002.4452 [hep-ph]].

\bibitem{Mann:2010jz}
  W.~A.~Mann, D.~Cherdack, W.~Musial and T.~Kafka,
  Phys.\ Rev.\  D {\bf 82}, 113010 (2010)
  [arXiv:1006.5720 [hep-ph]].

\bibitem{Kopp:2010qt}
  J.~Kopp, P.~A.~N.~Machado and S.~J.~Parke,
  Phys.\ Rev.\  D {\bf 82}, 113002 (2010)
  [arXiv:1009.0014 [hep-ph]].

\bibitem{Lee:1955vk}
  T.~D.~Lee and C.~N.~Yang,
  Phys.\ Rev.\  {\bf 98}, 1501 (1955).

\bibitem{Bell:1964ff}
  J.~S.~Bell and J.~K.~Perring,
  Phys.\ Rev.\ Lett.\  {\bf 13}, 348 (1964).

\bibitem{Bernstein:1964hh}
  J.~Bernstein, N.~Cabibbo and T.~D.~Lee,
  Phys.\ Lett.\  {\bf 12}, 146 (1964).

\bibitem{Barenboim:2009ts}
  G.~Barenboim and J.~D.~Lykken,
  Phys.\ Rev.\  D {\bf 80}, 113008 (2009)
  [arXiv:0908.2993 [hep-ph]].

\bibitem{Eotvos:1922pb}
  R.~V.~E\"{o}tv\"{o}s, D.~Pekar and E.~Fekete,
  Annalen Phys.\  {\bf 373}, 11 (1922).

\bibitem{Okun:1995dn}
  L.~Okun,
  Phys.\ Lett.\  B {\bf 382}, 389 (1996)
  [arXiv:hep-ph/9512436].

\bibitem{Dolgov:1999gk}
  A.~D.~Dolgov,
  Phys.\ Rept.\  {\bf 320}, 1 (1999).

\bibitem{JM}
  A.~S.~Joshipura and S.~Mohanty,
  Phys.\ Lett.\  B {\bf 584}, 103 (2004)
  [arXiv:hep-ph/0310210].

\bibitem{Grifols:2003gy}
  J.~A.~Grifols and E.~Masso,
  Phys.\ Lett.\  B {\bf 579}, 123 (2004)
  [arXiv:hep-ph/0311141].

\bibitem{GonzalezGarcia:2006vp}
  M.~C.~Gonzalez-Garcia, P.~C.~de Holanda, E.~Mass\'{o} and R.~Zukanovich Funchal,
  JCAP {\bf 0701}, 005 (2007)
  [arXiv:hep-ph/0609094].
In the conclusions of this paper, the quantities
  $\cos2\theta_{23}$ and $\sin2\theta_{23}$ should be replaced by $\cos^2\theta_{23}$ and
  $\sin^2\theta_{23}$, respectively.  With these modifications, the bounds
  derived in our work will be implied.

\bibitem{Bandyopadhyay:2006uh}
  A.~Bandyopadhyay, A.~Dighe and A.~S.~Joshipura,
  Phys.\ Rev.\  D {\bf 75}, 093005 (2007)
  [arXiv:hep-ph/0610263].  This paper gives a looser bound on $\alpha'$ than 
Ref.~\cite{GonzalezGarcia:2006vp} by a factor of 3/2.  

\bibitem{Samanta:2010zh}
  A.~Samanta,
  arXiv:1001.5344 [hep-ph].

\bibitem{gravtest}
E.~Fischbach and C.~L.~Talmadge, {\it The Search for Non-Newtonian Gravity}, Springer-Verlag, New York (1999);
  E.~G.~Adelberger, B.~R.~Heckel and A.~E.~Nelson,
  Ann.\ Rev.\ Nucl.\ Part.\ Sci.\  {\bf 53}, 77 (2003)
  [arXiv:hep-ph/0307284].

\bibitem{MINOSatm}
(MINOS results on atmospheric neutrinos and antineutrinos)
{\tt http://www-numi.fnal.gov/pr\_plots/index.html}

\bibitem{Wiebusch:2009jf}
  C.~Wiebusch for the IceCube Collaboration,
  arXiv:0907.2263 [astro-ph.IM].

\bibitem{Raby:2008pd}
  S.~Raby {\it et al.},
  arXiv:0810.4551 [hep-ph].
  For more physics cases relevant for DUSEL, see this Theory White Paper.

\bibitem{Ma:1997nq}
  E.~Ma,
  Phys.\ Lett.\  B {\bf 433}, 74 (1998)
  [arXiv:hep-ph/9709474].

\bibitem{Lee:2010hf}
  H.~S.~Lee and E.~Ma,
  Phys.\ Lett.\  B {\bf 688}, 319 (2010)
  [arXiv:1001.0768 [hep-ph]].

\bibitem{footnote}
We also note that neutrino decays occur in our scenario, $\nu_i \to \nu_j\, Z'$, 
with a decay rate $\Gamma(\nu_i \to \nu_j\, Z') \sim (\alpha'/8) (m_i^2 - m_j^2)^3/(m_i^3 m_{Z'}^2)$.
If applicable, cosmology \cite{Recombination} then requires $m_{Z'} \gsim 10^{-16}$~eV, 
eliminating the possibility of a long range $(1/R)$ Sun-Earth potential.  In that case, the Earth would 
have to provide the source for an observable $\nu_\mu-\nu_\tau$ potential difference in terrestrial 
experiments.  Such a change could be accommodated in our scenarios, but somewhat larger 
$\alpha'$ values would be needed for observable consequences in neutrino oscillations.
The details will be discussed in a forthcoming publication.

\bibitem{Recombination}
  S.~Hannestad and G.~Raffelt,
  Phys.\ Rev.\  D {\bf 72}, 103514 (2005)
  [arXiv:hep-ph/0509278];
  A.~Basboll, O.~E.~Bjaelde, S.~Hannestad and G.~G.~Raffelt,
  Phys.\ Rev.\  D {\bf 79}, 043512 (2009)
  [arXiv:0806.1735 [astro-ph]].

\bibitem{Friedland:2004ah}
  A.~Friedland, C.~Lunardini and M.~Maltoni,
  Phys.\ Rev.\  D {\bf 70}, 111301 (2004)
  [arXiv:hep-ph/0408264].
More detailed studies of the Super-Kamiokande
atmospheric data in G.~Mitsuka, Ph.D. Thesis, University of Tokyo, February 2009, arrive
at a similar constraint $\varepsilon_{\tau\tau}<0.15$, at 90\% confidence level.  See also,
  G.~Mitsuka  [Super-Kamiokande Collaboration],
  PoS {\bf NUFACT08}, 059 (2008).

\bibitem{Zeller:2001hh}
  G.~P.~Zeller {\it et al.}  [NuTeV Collaboration],
  Phys.\ Rev.\ Lett.\  {\bf 88}, 091802 (2002)
  [Erratum-ibid.\  {\bf 90}, 239902 (2003)]
  [arXiv:hep-ex/0110059].

\bibitem{Barr:2006it}
  G.~D.~Barr, T.~K.~Gaisser, S.~Robbins and T.~Stanev,
  Phys.\ Rev.\  D {\bf 74}, 094009 (2006)
  [arXiv:astro-ph/0611266].

\bibitem{FernandezMartinez:2010am}
  E.~Fernandez-Martinez, G.~Giordano, O.~Mena and I.~Mocioiu,
  Phys.\ Rev.\  D {\bf 82}, 093011 (2010)
  [arXiv:1008.4783 [hep-ph]].

\bibitem{DUSEL}
Talk by M. Diwan at
DURA Annual Meeting and DUSEL PDR Rollout, FermiLab, September 2-3, 2010.
({\tt http://www.dusel.org/workshops/fallworkshop10/\\milind\_dura\_ralk.pdf})

\bibitem{Friedman:1991dj}
  J.~A.~Frieman and B.~A.~Gradwohl,
  Phys.\ Rev.\ Lett.\  {\bf 67}, 2926 (1991).

\bibitem{Lee:2008pc}
  H.~S.~Lee,
  Phys.\ Lett.\  B {\bf 663}, 255 (2008)
  [arXiv:0802.0506 [hep-ph]].

\bibitem{Kaloper:2009nc}
  N.~Kaloper and A.~Padilla,
  JCAP {\bf 0910}, 023 (2009)
  [arXiv:0904.2394 [astro-ph.CO]].

\bibitem{Bernabei:2010mq}
  R.~Bernabei {\it et al.},
  Eur.\ Phys.\ J.\  C {\bf 67}, 39 (2010)
  [arXiv:1002.1028 [astro-ph.GA]].


\end{thebibliography}
\end{document}